# Direct measurement of surface interactions experienced by sticky microcapsules made from environmentally benign materials

*Hairou Yu[1] and Christopher L. Wirth[1]*

[1]Department of Chemical and Biomolecular Engineering, Case School of Engineering, Case Western Reserve University, Cleveland, Ohio 44106, United States

**Corresponding author**

Christopher L. Wirth

Chemical and Biomolecular Engineering Department

Case School of Engineering

Case Western Reserve University

Cleveland, OH 44106

wirth@case.edu

HY Orchid: 0009-0008-6510-6427

CLW Orchid: 0000-0003-3380-2029

ABSTRACT

We present a study combining experimental measurements, theoretical analysis, and simulations to investigate core-shell microcapsules interacting with a solid boundary, with a particular focus on understanding the short-range potential energy well arising from the tethered force. The microcapsules, fabricated using a Pickering emulsion template with a cinnamon oil core and calcium alginate shell, were characterized for size ($\sim 5 - 6$ μm in diameter) and surface charge ($\sim -20\ mV$). We employed total internal reflection microscopy and particle tracking to measure the microcapsule-boundary interactions and diffusion, from which potential energy and diffusivity profiles were derived. The potential energy profile was analyzed and simulated by considering electrostatic, gravitational, and tethered forces, while the diffusivity was compared to that of a solid particle-boundary interaction, inclusive of hydrodynamic forces. The diffusivity was represented as a normalized diffusion coefficient to eliminate the impact of fluid viscosity. The normalized diffusion coefficient of polymer-shell microcapsules ($\sim 0.02$) was found to be an order of magnitude smaller than that of solid polystyrene particles ($\sim 0.2$). The microcapsule sampled a potential well consisting of two distinct minima, as observed experimentally and supported by analytical expressions and Brownian dynamics simulations. A critical tethered height $h_{ct} = 49.8\ nm$ and the alginate radius of $r_g = 35.2\ nm$ were obtained from fitting our model to experimental data. This work concludes that these benign core shell microcapsules interact with a nearby boundary via a transient tethering interaction, overall producing a mild 'sticky' interaction that would likely be beneficial for applications in consumer products.

INTRODUCTION

Microcapsules (~ 1 μm – 10 μm in size) are prevalent in both natural and artificial environments. These colloids typically consist of two domains: a core that encloses materials and a shell that can function as a membrane for controlled release [1]. Microcapsules are designed for a variety of functions and applications and are deployed in a wide range of industries, including drug delivery [2-4], personal care [5], carbon capture [6], and the food industry [7-9]. Recent interest in manipulating the dynamic behavior of microcapsules has motivated a substantial amount of work [6, 10]. Notably, several excellent modeling studies have explored the elastic deformation of microcapsules, microcapsule stability in response to flow-induced shear stress, and microcapsule-microcapsule interactions [11-13]. These studies demonstrated the critical importance of ~kT scale surface interactions to microcapsules.

There are a variety of techniques capable of measuring the interaction between a micrometer scale particle and surface, including Atomic Force Microscopy (AFM), the Surface Force Apparatus (SFA), and Total Internal Reflection Microscopy (TIRM). AFM and SFA are robust techniques that have been used to measure surface forces in a variety of systems. For example, both techniques have been used to measure the classical theory of Derjaguin, Landau, Verwey and Overbeek (DLVO) forces [14-17]. For non-DLVO systems, pure attractive force such as the depletion force, pure repulsive force such as the steric force, and oscillatory forces have been studied [18-20]. Although both AFM and SFA have been adopted as useful techniques, one challenge that remains is conducting measurements with sufficient sensitivity to elucidate ~kT scale surface interactions. TIRM overcomes these challenges by directly measuring the Brownian fluctuations experienced by a colloidal particle separated a distance $h$ from a boundary and then subsequently inferring the potential energy well sampled by the particle from those fluctuations [21, 22]. The essence of TIRM is the collection of light scattered from a particle interacting with an evanescent field, which decreases exponentially in intensity with increasing $h$. The potential energy well is calculated by capturing many observations of height (~50,000 - 100,000 for an isotropic particle) and then assembling a potential energy well by assuming those sampled observations in height satisfy a Boltzmann distribution. This technique has successfully been applied to study the DLVO-forces experienced by a solid spherical particle [23, 24] and non-DLVO forces including steric, depletion, and tethering forces [25-27]. The dynamics of particles responding to external fields, for example A/C electric and magnetic fields, have also been studied and summarized [28-33]. Notably, TIRM is now being expanded to anisotropic particles by analyzing the morphology of scattering, rather than the integrated intensity as has been done until recently [34-36]. One benefit of analyzing the morphology as opposed to the integrated intensity of scattered light is the potential to sense changes in particle shape, which may serve as a tool to measure soft particle deformation. Previous work focused on measuring surface interactions of microcapsules with TIRM [37], but these studies focused on a subset of capsules that experience interactions with a single local potential energy minimum. Given that changes in the nature of the capsule itself and variations in the environment (ex. dispersed polymer, changes in pH, or flow) will all contribute to the interaction, there is a wide range of other phenomena experienced by microcapsules that are suitable to probe with TIRM.

The work described herein focuses on the fabrication and characterization of an oil filled microcapsule consisting of environmentally benign materials. The manuscript first describes the synthesis of oil-core capsules and subsequently characterizing the capsule. We then summarize a surprising observation of how these microcapsules interact with a nearby boundary as measured with TIRM. Raw data from TIRM measurements showed microcapsule fluctuations with significantly different temporal trends in height sampling as compared to a solid particle. Analysis of these data showed the microcapsule tended to sample a potential energy well with multiple local minima. These experimental data suggested the capsule experienced a surface interaction consisting of the typical DLVO interaction complemented by a transient tethering potential. We developed an analytical model based on this interpretation, as well as conducted Brownian dynamics simulations, that suggest microcapsules experienced a tethering or 'sticky' potential of interaction consistent with polymer-surface interactions. We further hypothesize the relative depth of these two minima depend on the local coverage of polymer in the vicinity of the capsule. Given the ubiquity of microcapsules in chemical products, cosmetics, and food, the design approach used for these benign capsules with a sticky interaction may find use in a variety of applications.

THEORY

**Conservative and non-conservative interactions experienced by a particle proximate to a boundary.** Consider the polystyrene particle and microcapsule shown in **Figure 1** close to a boundary.

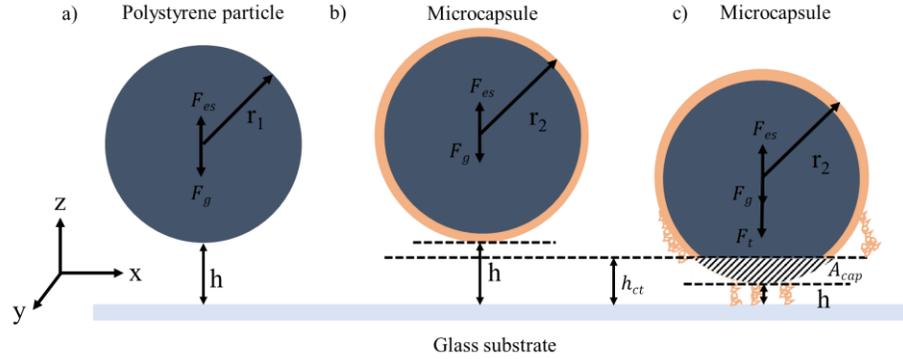

**Figure 1. Conservative interactions experienced by colloids near a boundary.** The variable $h_{ct}$ is the critical tethering height, below which the polymer chains are sufficiently long to bridge the gap $h$. **(a)** A polystyrene particle of radius $r_1$ separated a distance $h$ from a substrate. **(b)** An "untethered" microcapsule of radius $r_2$ with polymer layer separated a distance $h$ from a substrate such that $h > h_{ct}$. **(c)** A microcapsule of radius $r_2$ "tethered" with a boundary a distance $h$ from a substrate such that $h \leq h_{ct}$.

Forces experienced by colloids consist of both attractive and repulsive forces. Namely, repulsion arises from electrostatic double layer overlap $F_{es}(h)$, whereas the particle is attracted to the boundary via the net gravitational force $F_g$ and tethering force $F_t(h)$ [25]. We do not consider Van der Walls attraction. The overall force balance (see **eq. 1**) is:

$$F(h) = F_{es}(h) + F_g + F_t(h) \qquad (1)$$

where:

$$F_{es}(h) = \kappa B e^{(-\kappa h)} \quad (2)$$

$$\kappa = \sqrt{\frac{8\pi C e^2}{\varepsilon kT}} \quad (3)$$

$$B = 64\pi\varepsilon r \left(\frac{kT}{e}\right)^2 \tanh\left(\frac{e\Psi_1}{4kT}\right) \tanh\left(\frac{e\Psi_2}{4kT}\right) \quad (4)$$

The electrostatic repulsion depends on the dielectric permittivity of water $\varepsilon = \varepsilon_0 \varepsilon_r$, the elemental charge $e$, and the Stern potentials of the particle and the plate $\Psi_1$ and $\Psi_2$. Here, we assume the Stearn potentials are equal to the microcapsule and surface zeta potentials. $\kappa^{-1}$ is the Debye length, which depends on the ionic strength $C$, $k$ is the Boltzmann constant, and $T$ is temperature.

$$F_g = \frac{4}{3}\pi r^3 (\rho_p - \rho_s) g = G \quad (5)$$

The gravity force depends on the particle size (radius) $r$ and the density difference between particle $\rho_p$ and solution $\rho_s$. $g$ is the gravitational constant of $g = 9.81 m/s$.

$$F_t(h) = \theta \frac{3kTN}{2PL_T}(h - r_g) \quad (6)$$

$\theta$ is the probability of tethering, ranging from 0 (no tethering) to 1 (all chains tethered). $N$ is the number of chains capable of tethering when the microcapsule is below the tethering height $h_{ct}$. $P$ is the persistence length of the polymer chain, $L_T$ is the contour length of the polymer chain, and $r_g$ is the radius of gyration of the polymer chain.

For the most part, the expressions shown above have been applied widely to a variety of experimental systems [25]. However, one unique approach used herein is how we calculated the tethering force, which is based on Hooke's law (see **eq. 6**)) but implements a probability $\theta$ of tethering depending on the proximity of the particle to the boundary. There are two important features of how $\theta$ is implemented to our work. First, we assumed that tethering only occurs when the microcapsule fluctuates equal or below the tethering height $h \leq h_{ct}$ (see **Figs. 1(b) & 1(c)**), such that $\theta = 0$ when $h > h_{ct}$. The second important feature is that at heights below the tethering height, $h \leq h_{ct}$, the probability of tethering can vary between 0 and 1, $0 \leq \theta \leq 1$. The tethering probability then works as a pseudo measure of coverage when it multiplies the maximum available chain number $N$. The maximum available chain number was obtained by setting it equal to the quotient of the total tethered cap area ($A_{cap}$) and the area of a single chain ($A_{rg} \sim \pi r_g^2$): $N = \frac{A_{cap}}{A_{rg}}$.

Finally, we simplified the denominator of the tethering interaction based on the physical dimensions of the polymers we used, such that $PL_T = r_g^2$. Following simplification, the tethering force is:

$$F_t(h) = \theta \frac{3kTr}{r_g^4}(h_{ct} - h)(h - r_g) \tag{7}$$

The conservative interaction potential was illustrated by integrating the force balance and plotting the potential energy well (see **Fig. 2**) for values typical of colloidal systems; namely $r_g = 100\ nm$, $h_{ct} = 150\ nm$, $C = 10\ mM$, and $r = 2.5\ \mu m$. Two local minima appear when a colloid experiences tethering, whereas the typical DLVO like interaction (absent of tethering) displays a single minimum. For the tethered colloid, the local minimum at smaller separation distances occurs when there is a balance of electrostatic forces, tethering, and gravity, whereas the second minimum at a larger separation distance occurs when the microcapsule becomes untethered and the potential energy profile only depends on electrostatics and gravity. Note that in the case of a tethered colloid, the most probable height is at a slightly lower separation distance as compared to the case of the untethered colloid.

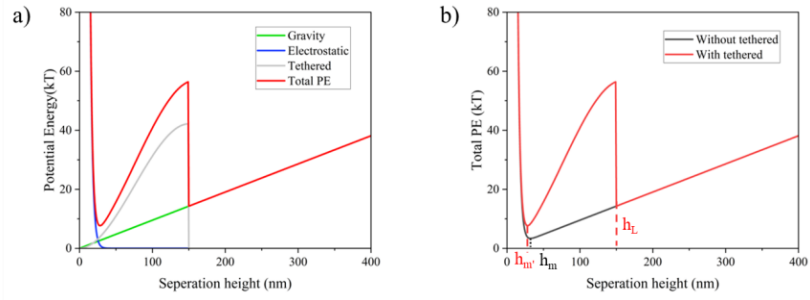

**Figure 2. Potential energy well with and without tethering interaction. (a)** Components of the total interaction energy for a tethered system. **(b)** Comparison between a tethered and untethered system. The potential energy well displays two local minima, with the relative separation distance between the two minima dependent on system properties.

Later in the manuscript, the experimental potential energy landscape will be presented as a function of the most probable height $h_m$, which is the height that is sampled most often by the colloid during an experiment and is different from the average height [38]. As has been done previously for a particle in the absence of tethering, the potential energy as a function of relative height $h - h_m$ can be presented as [21, 22]:

$$\phi(h) - \phi(h_m) = \frac{G}{\kappa}(e^{-\kappa(h-h_m)} - 1) + G(h - h_m) \qquad (8)$$

and scaling with the thermal energy *kT*:

$$\frac{\phi(h) - \phi(hm)}{kT} = \frac{1}{kT}[\frac{G}{\kappa}(e^{-\kappa(h-h_m)} - 1) + G(h - h_m)] \qquad (9)$$

These established expressions become more complicated for the microcapsule system of concern herein. One needs to account for the possibility of multiple minima, here represented as $h_{ct}$ and $h_{m'}$, which are the tethering height and most probable height specifically for a tethered colloid. Following a similar procedure to how **equations 8 & 9** were obtained, the analytical expression can be presented as follow, with detailed derivation in the support information (SI). For $I < I_{h_{ct}}$ ($h > h_{ct}$), there is no tethered force, and the potential energy is expressed as:

$$\frac{\phi(h) - \phi(h_{m'})}{kT} = \frac{1}{kT}[B(e^{-\kappa h_{m'}}(e^{-\kappa(h-h_{m'})} - 1) + G(h - h_{m'})] - \theta \frac{3r}{2r_g^4}(h_{ct} - h_{m'})(h_{m'} - r_g)^2$$

(10)

For $I \geq I_{h_{ct}}$ ($h \leq h_{ct}$), the tethering force is operating, and the potential energy is expressed as:

$$\frac{\phi(h) - \phi(h_{m'})}{kT} = \frac{1}{kT}[B(e^{-\kappa h_{m'}}(e^{-\kappa(h-h_{m'})} - 1) + G(h - h_{m'})] + \theta \frac{3r}{2r_g^4}[(h_{ct} - h)(h - r_g)^2 - (h_{ct} - h_{m'})(h_{m'} - r_g)^2]$$

(11)

The colloid will sample the potential energy profile because of fluctuations in position, which are damped by the viscous fluid, leading to non-conservative forces contributing to the dynamics of the colloid. The nearby boundary will further dampen these fluctuations such the translational diffusion coefficients in both the direction normal $D_{h\perp} = f_{h\perp} \frac{kT}{6\pi\eta r}$ and parallel $D_{h\parallel} = f_{h\parallel} \frac{kT}{6\pi\eta r}$ to the boundary will decrease [39]. For a solid colloidal particle in a viscous fluid moving slowly towards or parallel the boundary, the correction parameters $f_{h\parallel}$ and $f_{h\perp}$ are known and have been widely applied throughout the last few decades [40-43]. Herein, we use these parameters to correct the diffusion coefficients employed in our Brownian dynamics simulations described later in the manuscript. These correction factors are: [40]

$$f_{h\perp} = \frac{6\left(\frac{h}{r}\right)^2 + 2\left(\frac{h}{r}\right)}{6\left(\frac{h}{r}\right)^2 + 9\left(\frac{h}{r}\right) + 2} \tag{12}$$

$$f_{h\parallel} = \frac{12420\left(\frac{h}{r}\right)^2 + 5654\left(\frac{h}{r}\right) + 100}{12420\left(\frac{h}{r}\right)^2 + 12233\left(\frac{h}{r}\right) + 431} \tag{13}$$

Note that there are additional effects on the diffusion coefficient for a colloid near a boundary with an adsorbed polymer layer. There is theoretical and experimental work revealing the effect of lubrication between polymer layers, derived correction parameters for uncompressed polymer layer and compressed polymer layer with homogeneously increased density [44, 45]. Given the some of the physical uncertainties of our system, we chose not to utilize such expressions and rather use hindrance expressions for a solid sphere in a viscous liquid.

**Simulating the Brownian dynamics of a microcapsule.** Brownian dynamics simulations were conducted for a microcapsule fluctuating in the z-axis normal to the nearby boundary and experiencing conservative and non-conservative interactions as described above. The algorithm for these simulations is based on a previous paper and executed in MATLAB [46-48]. Briefly, a stepping algorithm was obtained by Ermak and McCammon [49], via solution of a Langevin equation. Namely, the height of the particle can be predicted as:

$$h_{t+\triangle t} = h_t + \frac{dD_{h\perp}}{dh} \triangle t + \frac{D_{h\perp}}{kT} F_h \triangle t + H_{\triangle t} \qquad (14).$$

In the algorithm, the $h_{t+\triangle t}$ is the calculated height based on the previous height $h_t$ with the influence of the net conservative force $F_h$, diffusion coefficient $D_{h\perp}$ hindered by the nearby boundary, and a random height displacement $H_{\triangle t}$. The random height displacement is calculated such that it is chosen from a Gaussian distribution that satisfies the $<H> = 0$ and the variance is $<H>^2 = 2D_{h\perp} \triangle t$.

Results from this stepping algorithm were obtained for a colloid at an initial height of 50 nm with a time step Δt = 0.2 ms and observations >60,000. The time step was selected based on the control simulations of polystyrene particles, which only consider the gravity and electrostatic forces. The control simulation results were compared with the analytical results and 0.2 ms was selected as an appropriate time step in following simulations. Detailed information can be found in SI. A series of heights were predicted, which then allowed for calculation of integrated intensity (as obtained in a TIRM measurement) and ultimately the potential energy profile sampled by the colloid. As will be described in greater detail later in the manuscript, these data were particularly helpful in not only comparing with the dynamics measured in experiments, but also the computed potential energy landscape. The parameters used in the simulations presented herein are showing in **Table 1**.

| Parameter name | Symbol | | Unit |
| --- | --- | --- | --- |
| Salt concentration* | $C$ | fit | mmol/L |
| Temperature | $T$ | 298 | K |
| Particle radius | $r$ | 0.000003 | m |
| Dielectric constant | $\varepsilon_r$ | 78.5 | |
| Dielectric permittivity of vacuum | $\varepsilon_0$ | 8.85 x $10^{-12}$ | |
| Density of particle | $\rho_p$ | 1611 | kg/m$^3$ |
| Density of solution | $\rho_s$ | 1000 | kg/m$^3$ |
| Viscosity of water | $\eta$ | 0.000887 | N s/m$^2$ |
| Viscosity of complex solution | $\eta_2$ | 0.004017 | N s/m$^2$ |
| Zeta potential particle | $\Psi_1$ | -0.02 | V |
| Zeta potential wall | $\Psi_2$ | -0.05 | V |
| Anion charge | $z_A$ | -1 | |
| Cation charge | $z_C$ | 1 | |
| Initial height | $h_0$ | 50 | nm |
| Time step | $\Delta t$ | 0.0002 | s |
| Radius of alginate* | $r_g$ | fit | nm |
| Critical tethered height* | $h_{ct}$ | fit | nm |
| Possibility of tethering* | $\theta$ | fit | |

**Table 1.** Parameters used to simulate the hindered Brownian motion. *Fitted parameters from analytical expressions.

MATERIAL AND METHODS

**Materials and Microcapsule Fabrication.** Low dispersity polystyrene (PS) particles with 5 μm diameter (Lot #BCCG3410), cinnamon leaf oil, acetic acid (>99.7%), sodium alginate, and sodium chloride (>99.6%) were bought from Sigma Aldrich (USA). The $CaCO_3$ nanoparticles, ranging in nominal diameter between 15 nm - 40 nm were purchased from SkySpring Nanomaterials, USA. Fluorescent tag, Nile Red (Ex: 552nm, Em: 636nm), was purchased from Invitrogen, Thermo Fisher Scientific, USA. All water used in this work was ultra-pure water (18.2 MΩ) generated from Arium Mini Ultrapure Water Systems (Sartorius, USA). All chemicals were used as received.

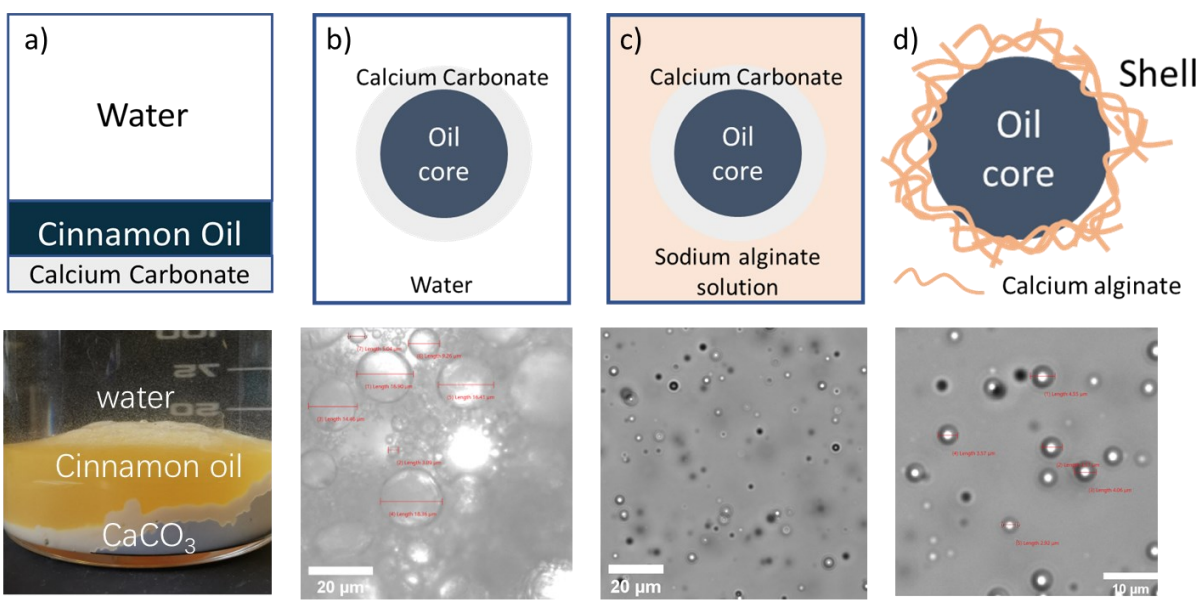

**Figure 3.** Illustrations and images of microcapsule fabrication processes. **(a)** Mix raw material with a homogenizer to form **(b)** oil-in-water Pickering emulsion and then relaxed for phase separation. **(c)** Redisperse the emulsion into sodium alginate solution. Adjust the pH with acetic acid to release $Ca^{2+}$ ions to form **(d)** a polymer shell.

The fabrication route for microcapsules was similar to one that had been previously published (**Fig. 3**) [50]. Briefly, the process started by first preparing a $CaCO_3$ nanoparticle dispersion with a high-speed homogenizer (IKA, model Ultra-Turrax T-25, Germany) at 10,000 rpm for 10 minutes. Next, 30 mL of cinnamon oil was emulsified in 70 mL of the 5 wt.% $CaCO_3$ nanoparticle dispersion by the high-speed homogenizer at 5000 rpm for 30 minutes. Those emulsions were then allowed to relax for 24 hours. Sodium alginate solution in water then prepared by stirring, with 5 mL of the emulsion redispersed into 100 mL of 2.5wt% sodium alginate solution via stirring for 10 minutes. The complex fluid then consisted of an oil-in-water emulsion with sodium alginate in the continuous phase. The microcapsule shell was cross-linked by adding 1M acetic acid drop to drop the pH = 4. The pH of the complex fluid was monitored following addition of the acetic acid, with a series of fabrication processes conducted at systematically varied reaction time. The total reaction process time was varied for 10 minutes, 60 minutes, 4 hours, and 24 hours. The reaction was then stopped by adding 200 mL ultra-pure water. The capsule solution was then filtered with filter paper (Pore size 100 μm, Whatman, Germany). The fabrication processes were conducted under room temperature.

Microcapsules were characterized in a variety of ways to approximate their size, charge, core size, and shell thickness. The size distribution and electrophoretic mobility were measured with a Zetasizer Nano ZS (Malvern). In addition, scanning electron microscopy (SEM) measurements were taken to measure the shape and the size of the microcapsules in vacuum at 15kV. The core size was approximated by adding fluorescent tag Nile red to the oil phase. The sizes of the oil-core and entire microcapsule were captured and measured via optical microscope under the fluorescent and bright field to calculate the thickness of the calcium alginate shell. Three repeats were taken for each sample. All measurements were taken under the original solution at the room temperature.

**Total Internal Reflection Microscopy (TIRM)**. TIRM was used to measure the Brownian motion of microcapsule close to the glass boundary and infer the potential energy landscape. TIRM has been conducted extensively over the past three decades, being described in several previous papers [21, 22]. The system used herein is based on an upright microscope system (Olympus BX51WI) (see **Figs. 4(a) & 4(c)**), including a 633 nm laser (PLM-633-PMF, NECSEL) which passes through a single-mode optical fiber (NECSEL) and couples to an aspherical collimator (CFC-8X-A, f = 7.5 mm, Thorlabs). The collimator is directly attached to a custom-built mount that holds a linear polarizer (p-polarization, FBR-LPVIS 500–720 nm, Thorlabs) pointed to a custom-made trapezoidal prism (BK7 glass, width: 25 mm, length: 50 mm at the larger facet) with diagonal sides of 68° and 75°. For all experiments in this paper, the angle of incidence was 75°, which is larger than the critical angle ($\theta_c = 61°$) required to achieve total internal reflection in a glass–water ($n_1 = 1.515$, $n_2 = 1.33$) interface. This arrangement leads to an evanescent decay length of $\beta^{-1} = 108$ nm. The laser power was controlled by software (Raman Boxx, NECSEL) from 0.1 mV to 5 mV.

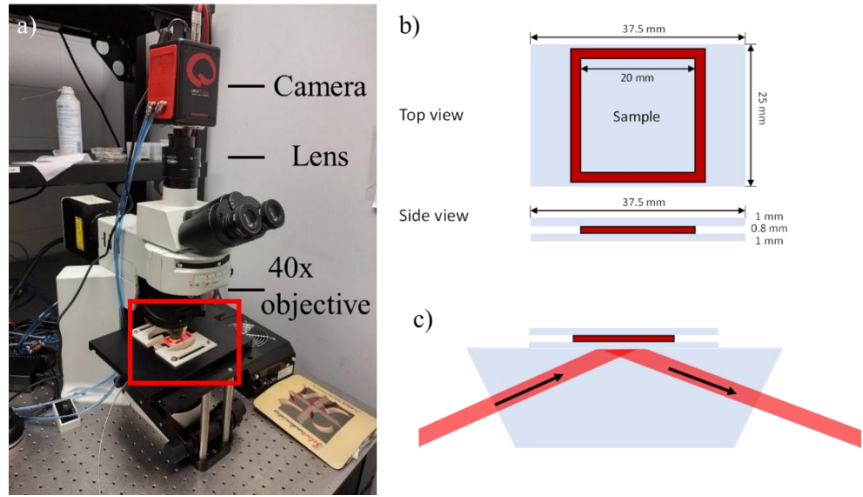

**Figure 4.** Image and schematics of TIRM setup. **(a)** The laser and prism are held on a customized mount. **(b)** Sample is placed in a sandwiched fluid cell which is made from microslides and gasket.

**(c)** The sample is placed on the prism by immerse oil to reach the reflective index match.

The fluid cell was placed on top of the prism using immersed oil (Resolve, Thermo Scientific, n = 1.515) to match the refractive index with glass. The structure of the fluid cells for TIRM experiments are shown in **Figure 4(b)**. The enclosed fluid cells were made up of two microslides sandwiched a gasket and the dimensions were shown. The microscope slides were sonicated in isopropyl alcohol and acetone for 15 minutes each and then cleaned with the air Plasma cleaner (Tergeo, PIE Scientific) for 2 minutes. Scattered light was captured by digital CMOS camera (ORCA-Fusion, C14440-20UP, Hamamatsu). The region of interest was set to 200 x 200-pixel size and the fluctuations of scattered light was captured for 20 minutes at a frame rate of 50 frames per second (fps). These conditions produced image stacks of 60,000 observations, which were saved as tiff files and subsequently analyzed via a custom MATLAB code to obtain the temporal intensity fluctuations and potential energy profile. We used the established procedure for converting the observations of integrated intensity to a potential energy profile [22].

**Particle tracking analysis to measure lateral diffusion**. As described above, TIRM was used to measure fluctuations normal to the boundary. In addition, we aimed to track fluctuation information lateral to the boundary via the diffusion coefficient in the x-y plane $D_{measured}$. The lateral diffusion coefficient was measured by tracking the x-y positions of either microcapsules or particles and subsequently calculating the mean square displacement [51]. Each colloid has a strong scattering light signal from TIRM, the center of which was tracked by a custom MATLAB code (see **Fig. 5(a)**). The whole TIRM image stack, consisting of 60,000 frames, was cut into intervals of 1000 frames, over which the mean squared displacement was calculated for each interval. The slope of the mean square displacement is equal to the product of four and the lateral diffusion coefficient ($m = 4D$) as shown in **Figure 5(b)**. Control experiments were conducted by measuring the diffusion coefficient of polystyrene particles in a salt solution.

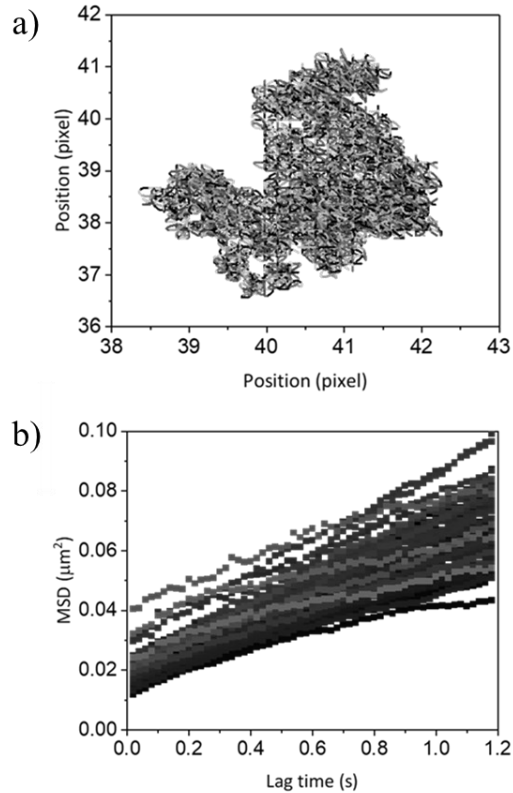

**Figure 5.** Polystyrene particle undergoes Brownian motion close to the boundary. **(a)** Position of the particle fluctuating and **(b)** MSD in x-y plane. The gray scale intensity of the position represents the height of the particle. The lighter gray scales represent higher separation distance of the particle away from the boundary.

An important aspect of the analysis described herein is normalization of the diffusion coefficient by the bulk diffusion coefficient after correcting for the static error (see **Eq. 15**). As will be described in more detail later in the manuscript, we do this primarily to evaluate whether there is an additional lateral hindrance to motion of the microcapsule beyond typical hydrodynamic hindrance. The bulk diffusion coefficient was calculated from the Stokes-Einstein equation, using the viscosity of fluid accounting for the presence of polymer when applicable (see **Eq. 16**). The normalized diffusion coefficient is then defined as:

$$D_N = \frac{D_{measured} - D_{error}}{D_{bulk}} \tag{15}$$

$$\eta_p = \frac{kT}{D_{bulk} 6\pi r} \tag{16}$$

where, $\eta_p$ is the viscosity of the polymer solution and $D_{bulk}$ is the diffusion coefficient of polystyrene particle in polymer solution.

RESULTS AND DISCUSSIONS

**Microcapsule characterization**. Initially, SEM imaging and sizing of microcapsules that had been fabricated with a one-hour gelation time was conducted (see **Fig. 6**). SEM imaging suggests the microcapsules have a spherical morphology with the center collapsed, indicating a core-shell structure. Further, dynamic light scattering of the microcapsule suspension revealed a multimodal size distribution (see **Fig. 6(b)**). These data suggest in addition to the microcapsules (~ 5000 nm) suspended in solution, there are also alginate micellar aggregates [7] (~ 400 nm) and $CaCO_3$ nanoparticles (~ 45 nm). The core-shell structure suggested by the SEM imaging was then confirmed by measuring the diameters of the inner core and outer shell using fluorescent microscopy (see **Fig. S2**). Further, we evaluated the impact of gelation time by fabricating microcapsules with systematic variation in the time they were exposed to acidic pH. Results from measurements in which the microcapsules had gelation times from 10 minutes to 24 hours suggest gelation time had a minor at best impact on the capsule wall thickness (see **Table 2**). These data suggest the hydrodynamic size of ~5 μm and surface charge of ~-20 mV remained independent of gelation time. The microcapsules discussed later in this manuscript were fabricated with a gelation time of four hours.

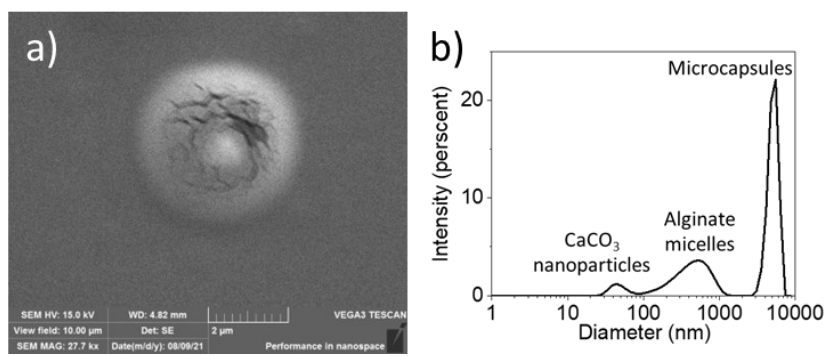

**Figure 6.** Microcapsule synthesized after one hour gelation time. **(a)** SEM image and **(b)** DLS size distribution. Three peaks from low

to high are 45 nm, 400 nm, and 5000 nm, corresponding to $CaCO_3$ nanoparticles, alginate micelles, and microcapsules.

| Gelation time | Zeta-potential [mV] | Hydrodynamic diameter [$\mu m$] |
|---|---|---|
| 10 min | $-21.5 \pm 1.08$ | $4.30 \pm 0.12$ |
| 1h | $-21.5 \pm 1.79$ | $4.82 \pm 0.36$ |
| 4h | $-21.3 \pm 1.11$ | $5.18 \pm 0.15$ |
| 24h | $-22.9 \pm 2.32$ | $5.25 \pm 0.30$ |

**Table 2.** DLS size and zeta-potential of microcapsules fabricated across varied gelation time. Both the size of ~5 μm and the surface charge of ~-20mV are independent of gelation time.

**Total Internal Reflection Microscopy (TIRM) of a microcapsule.** Once fabricated, we tracked the fluctuations of microcapsules near a boundary using TIRM. These temporal fluctuations were then compared to control experiments with rigid polystyrene spheres of a similar size in a low concentration salt solution. Notably, these systems had significantly different environments as revealed through optical microscopy (see **Figs. 7(a) & 7(c)**), suggesting that some polymer remained in the microcapsule system, despite using a filtration process. Further, the initial intensity fluctuations for the two systems have significant differences in their signatures (see **Figs. 7(b) & 7(d)**).

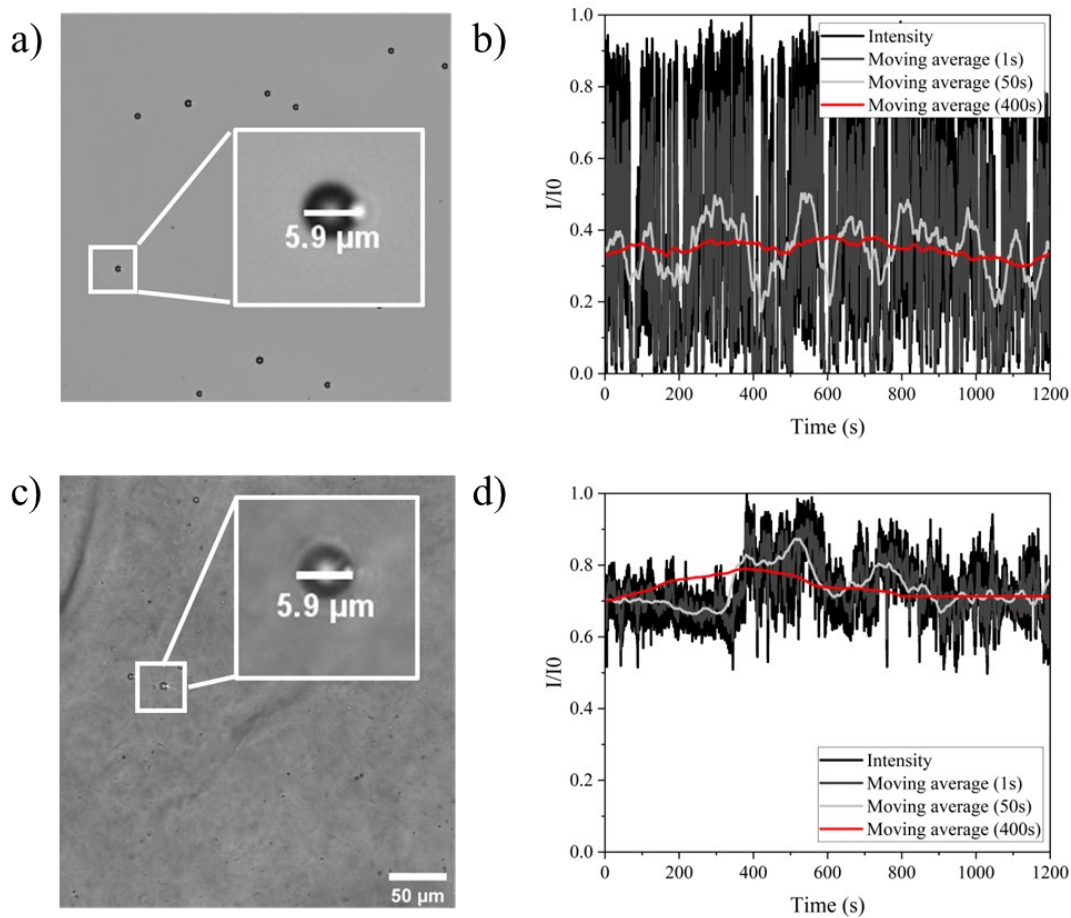

**Figure 7.** Optical microscope images and intensity profiles for **(a, b)** polystyrene particles and **(c, d)** microcapsules. Moving average lines for three interval time from low to high are plotted. The low interval time (1 s) is in dark gray, the medium interval time (50 s) is in light gray, and the long interval time (400 s) is in red. Note the intensity fluctuations from the microcapsule experience a long time (~100s excursion) indicated by the red moving average.

Although both have temporal fluctuations in intensity that are stochastic, both the range over which intensity fluctuations occur and the moving average suggest a difference in the dynamic behaviors of the two systems. Recall the intensity is exponentially dependent on the separation distance of the colloid from the nearby boundary, with a larger scattering intensity occurring at smaller separation distances. Both temporal scans of intensity are stochastic because of both colloids, the polystyrene bead and microcapsule, experiencing Brownian fluctuations, which allows the colloids to sample their respective potential energy profile. Averaging of these fluctuations over a long period of time as compared to the sampling rate should produce a single average scattering intensity, which corresponds to the average separation distance that is consistent with a potential energy landscape experienced by the particle that does not depend on time. Note the data reported in **Figure 7**: the high interval time (400 s) moving average of the intensity fluctuations for the polystyrene bead does not itself appear to have a temporal dependence, whereas there are significant excursions in the moving average of intensity fluctuations produced by the microcapsule. Excursions in the moving average of scattering suggest the microcapsule experienced an interaction that was variable during the experiment. These temporal shifts of moving average in the intensity fluctuations were observed in multiple microcapsule experiments and suggest the microcapsule system is experiencing additional interactions beyond electrostatic repulsion and gravity, with that additional interaction dynamic in nature.

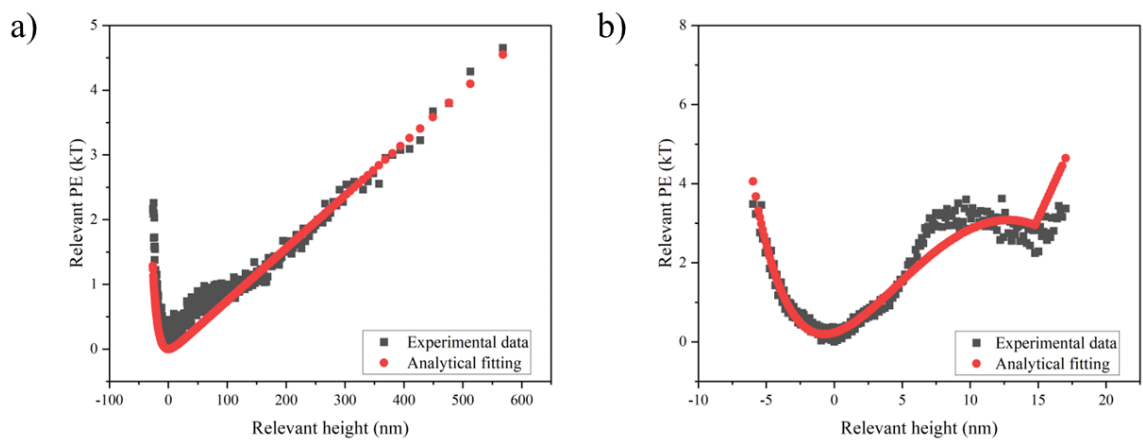

**Figure 8.** Relevant potential energy as a function of relevant height for **a)** polystyrene particle in salt solution and **b)** microcapsule in polymer solution. Pearson correlation coefficient for polystyrene system is 0.99 and for microcapsule system is 0.935. Both figures show good agreement between experimental data and analytical fitting.

Intensity fluctuations were analyzed to obtain the potential energy profiles for both the polystyrene particle and microcapsule (see **Fig. 8(a) & 8(b)**). As expected from the intensity fluctuations, the control system experienced a DLVO potential energy profile consisting of electrostatic repulsion exponentially dependent on separation distance and gravity linearly dependent on separation distance. Summed together, these interactions form an expo-linear potential energy profile with a single minimum corresponding to the most probable separation distance. The potential energy profile for the microcapsule was dramatically different in that there were two local minimums. Multiple local minimums arise because of the temporal fluctuations of intensity shown in **Figure 9**. Once a histogram is assembled from these intensity fluctuations, multiple maxima appear for the microcapsule that translate to multiple minima in the potential energy landscape (See **Fig. 9(a)**). We hypothesized this difference arose from a tethering interaction mediated by the polymer shell of the microcapsule, along with polymer that is potentially adsorbed to the nearby boundary. Similar interactions have been measured in systems of protein interacting with a polymer brush coated boundary [25, 52-54]. As described above, we theorized the tethering interaction was transient or 'sticky', in the sense that at certain points in the experiment, when the colloid was proximate to the boundary at $h \leq h_{ct}$, the microcapsule would be tethered to the nearby boundary with some probability θ that is a measure of the polymer coverage. These instances would correspond to the moving average intensity excursions to larger values (see **Fig. 7**). When at $h > h_{ct}$, the microcapsule would be untethered, experiencing an interaction akin to DLVO.

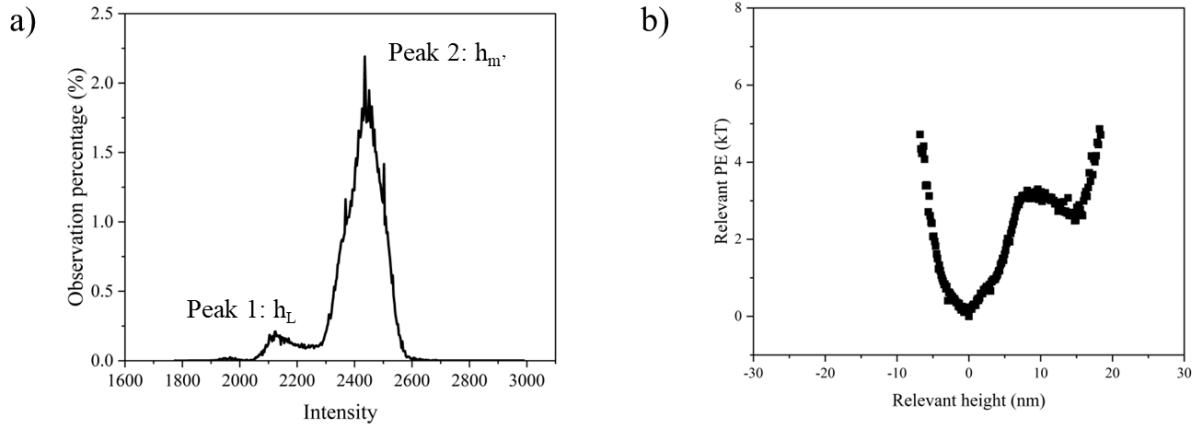

**Figure 9.** TIRM measurement of microcapsule fluctuating close to the boundary. **(a)** The intensity histogram shows two peaks with located at $h_{ct}$ and $h_{m'}$. **(b)** The measurement relevant potential energy profile. The x-axis is the relevant height compared to the $h_{m'}$, and the y-axis is the relevant height compared to the PE at $h_{m'}$.

| | | | |
|---|---|---|---|
| Salt concentration | $C$ | 8.3 | mmol/L |
| Possibility of tethering | $\theta$ | 0.9475 | |
| Radius of alginate | $r_g$ | 35.2 | nm |
| Critical tethered height | $h_{ct}$ | 49.8 | nm |

**Table 3.** Analytical fitted results for microcapsule.

We tested this hypothesis by fitting the developed model (see **eqs. 10 & 11**) to the experimental potential energy profile (see **Fig. 9(b)**) and the results are shown in **Table 3**. The models included both assumed ($B$, $r$) and fit ($\theta$, $C$, $h_{ct}$, $r_g$) parameters. Assumed radius $r = 3$ μm is based on the optical image obtained from the optical microscope. $B = 2.40 \times 10^{-17}$ is calculated and assumed based on the zeta potential obtained for microcapsule and glass. We need initial guess for $C$ and $h_{ct}$ at time = 0 s. Note that although we are unable to prepare the microcapsule samples with a known salt concentration because of the fabrication procedure, we were able to calculate an approximated ionic strength of C = 7.8 - 8.3 mM by running multiple control experiments with polystyrene beads in the same continuous phase, measuring the potential energy well, and obtaining the salt concentration from the Debye length. The initial guess of tethering height $h_{ct}$ at time zero was based on the maximum extended polymer length for our particular salt concentration [55] of C = 8 mM, such that $h_{ct} = 100\ nm$. The analytical fittings for both systems agree with the experimental data well (see **Fig. 8**) with a Pearson correlation coefficient of 0.99 (polystyrene system) and 0.935 (microcapsule system). The fitted results are shown in **Table 3**. The fitted $C$ agrees with the approximated ionic strength, $C_{fitted} = 8.3\ mM$ and the κ can be calculated κ $= 3.0 \times 10^8$.

**Lateral diffusion of a microcapsule**. The lateral diffusion coefficients of the control and microcapsule systems were measured to further probe these phenomena. The premise of these experiments was that a fluctuating, yet tethered, microcapsule would experience a lateral diffusion coefficient strongly hindered beyond the hydrodynamic hindrance from the nearby boundary. The diffusion coefficients parallel to the substate were calculated from mean square displacement with a particle tracking method described earlier in the manuscript. The diffusion coefficients were normalized with bulk diffusion coefficients to eliminate the impact of viscosity on the diffusion coefficients and isolate the impact of hydrodynamic hindrance. The viscosity of the polymer solution was found to be approximately four times higher than the viscosity of the salt solution (~ 4.017 cP). The normalized diffusion coefficients were calculated and compared between polystyrene particles and microcapsules in salt and polymer solutions as shown in **Figure 10**.

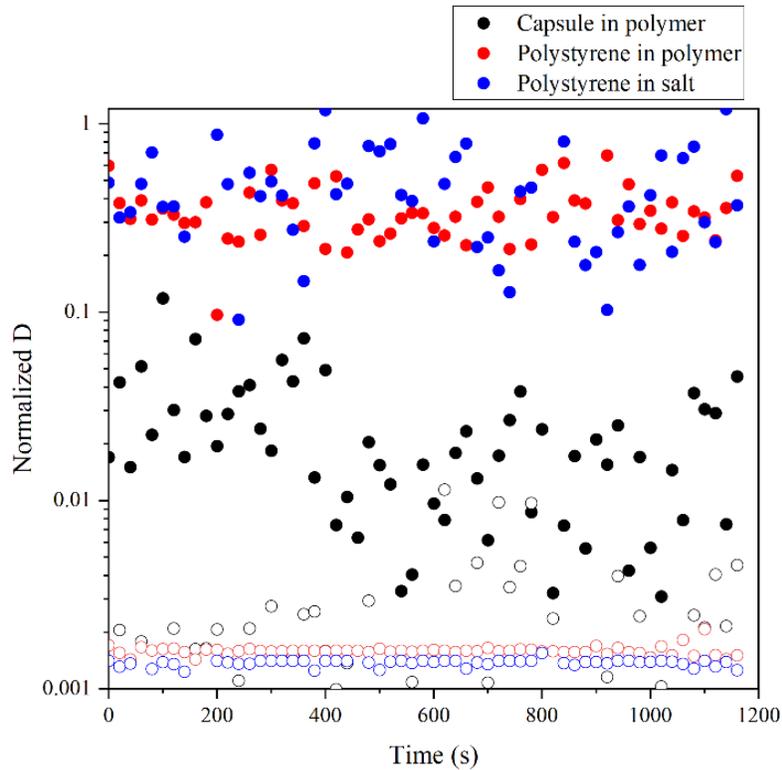

**Figure 10.** Normalized diffusion coefficients of microcapsule in polymer solvent (black), polystyrene particle in polymer solvent (red), and polystyrene particle in salt solvent (blue). The solid dots are the diffusive diffusion coefficients, and the open dots are adhesive diffusion coefficients.

The collection of data summarized in **Figure 10** has several important messages. First, note that all normalized diffusion coefficients were < 1. This was unsurprising, in that the diffusion coefficient for all colloids neighboring a boundary should be smaller as compared to the bulk value of diffusion coefficient. Notably, the normalized diffusion coefficients of the polystyrene bead in both the salt and polymer solutions were approximately the same with a value of ~ 0.2. As noted above, the diffusion coefficient was corrected for the difference in viscosity of these two fluids such that the hydrodynamic hindrance was isolated in the results. These data from the polystyrene particle suggests a similar hindered behavior for the polystyrene particle in a viscous fluid regardless of the viscosity of that fluid.

However, the normalized diffusion coefficients for the microcapsule were significantly different from the polystyrene particles in either the salt or polymer solutions. We found the microcapsule to have a normalized diffusion coefficient of ~ 0.02, approximately one order of magnitude smaller than for the polystyrene particles. Further, we found the normalized diffusion coefficient of the microcapsule to be about one order of magnitude larger than an adhered colloid (~0.002), including other completely stuck microcapsules. These data suggest that although the microcapsule was strongly hindered, it still behaved as a Brownian colloid to the extent it fluctuated far more than its adhered counterpart. The strong hindrance tended to be transient, with the normalized diffusion coefficient being larger earlier in the experiment as compared to later (see **Fig. 10**). From these data, it appears the microcapsule is experiencing a strong restriction in lateral movement during portions of the experiment. Together with the potential energy profiles summarized earlier, the microcapsule appears to interact with the neighboring boundary via a sticky interaction mediated by a transient tethering phenomenon.

**Simulated Brownian fluctuations of a microcapsule normal to the boundary.** Fitting parameters ($\theta$, $C$, $h_{ct}$, $r_g$) obtained from the analytical fitting of experimental data, along with the force balance described above (see **Eqs. (1) – (7)**), were used to conduct Brownian dynamics simulation of the motion of a microcapsule normal to a boundary. These data were then used to assembly a potential energy profile (see **Fig. 11**). As was observed in the experiment, simulated separation height from a microcapsule had transient changes in the moving average, suggesting a dynamic interaction that changed over the time scale of the experiment.

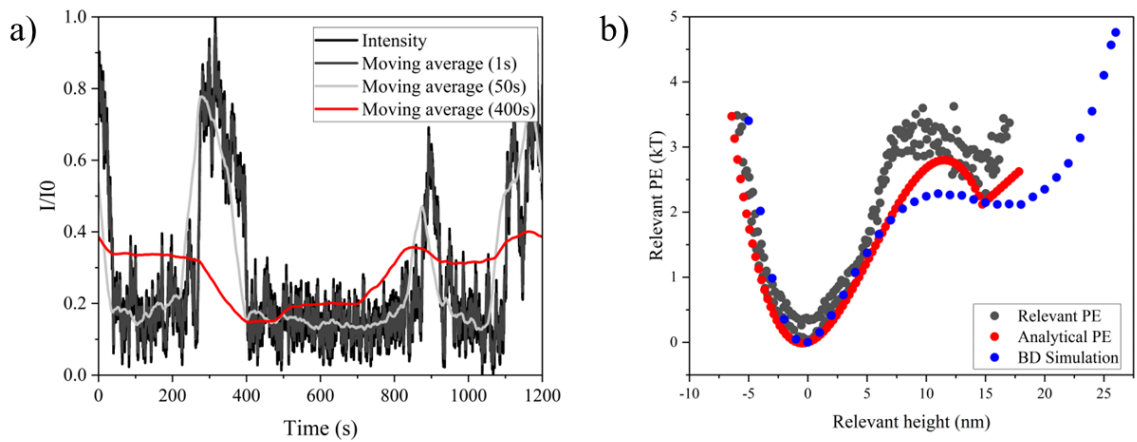

**Figure 11.** Brownian dynamic simulation of **(a)** fluctuation height and **(b)** potential energy profile using the parameters from analytical fitting parameters. Shift fluctuation positions were observed under different time scales. The similarity of potential energy profile from experimental data (black), analytical fitting (red) and Brownian dynamic simulation (blue) indicates the feasibility of new model.

Assembling of these data into a potential energy profile further supports the picture of a dynamic tethering interaction. The potential energy of the simulated microcapsule appears to have a weak second minimum matching that of the experimental and analytical profile. One potential origin of the lack of agreement is in the choice of the hindrance for the normal diffusion coefficient of the microcapsule. Previous works have shown that it is important to apply correction parameters for both uncompressed and compressed polymer layers when calculating the near boundary diffusivity [44, 54]. It has shown that the polymer layer thickness and polymer mesh size play a role in the polymer-boundary system [45].

CONCLUSIONS

This manuscript summarizes a comprehensive study combining experiments, theory, and simulations to explore the interactions of 'sticky' microcapsules fabricated from environmentally benign materials. The facile fabrication method produced microcapsules ~ 5 - 6 μm in diameter, with a wall thickness that remained largely independent of fabrication time. Measurements to track the fluctuations of these microcapsules, both normal and parallel to a nearby boundary, suggest that these colloids experienced a transient attractive interaction that intermittently tended to turn 'on' or 'off' throughout a long experiment. We developed a model that treated this interaction as a transient tethering interaction, which becomes relevant only when the microcapsule separation distance is equal or less than a critical tethering height $h_{ct}$, determined by the physical dimensions of the polymer. This analytical model fit our experimental data well and was further tested with a simulation tool that incorporated microcapsule dynamics. Although the agreement between the experimental and simulation results was good, there are features suggesting that a more detailed treatment of the normal diffusion coefficient is needed. From this work, we conclude that these benign, core shell microcapsules interact with a nearby boundary via a transient tethering interaction, resulting in a mild 'sticky' interaction behavior that would likely be advantageous for applications in consumer products. Further work will aim to understand the influence of these polymers on the diffusion coefficient normal to the boundary.

**Acknowledgements**
This work was supported by the National Science Foundation CAREER Award NSF No. 2023525 and NSF No. 2314405.

# SUPPORTING INFORMATION

# Direct measurement of surface interactions experienced by sticky microcapsules made from environmentally benign materials

*Hairou Yu[1] and Christopher L. Wirth[1]*

[1]Department of Chemical and Biomolecular Engineering, Case School of Engineering, Case Western Reserve University, Cleveland, Ohio 44106, United States

**Corresponding author**

Christopher L. Wirth

Chemical and Biomolecular Engineering Department

Case School of Engineering

Case Western Reserve University

Cleveland, OH 44106

wirth@case.edu

HY Orchid: 0009-0008-6510-6427

CLW Orchid: 0000-0003-3380-2029

**S1. Derivation of the analytical expression.** Based on the description in the manuscript (eq. 1), the analytical expression for microcapsule-boundary interactions can be derived as follows:

For $h > h_{ct}$, tethered force is not applied. The overall force balance is:

$$F(h) = F_{es}(h) + F_g = \kappa B e^{(-\kappa h)} + \frac{4}{3}\pi r^3 (\rho_p - \rho_s)g = \kappa B e^{(-\kappa h)} + G \quad (S1)$$

And the potential energy of the microcapsule located at $h > h_{ct}$ can be integrated as:

$$\phi(h) = Be^{-\kappa h} + Gh \quad (S2)$$

For $h \leq h_{ct}$, tethered force is considered. The overall force balance is:

$$F(h) = F_{es}(h) + F_g + F_t = \kappa B e^{(-\kappa h)} + \frac{4}{3}\pi r^3 (\rho_p - \rho_s)g + \theta \frac{3kTr}{r_g^4}(h_{ct} - h)(h - r_g) =$$

$$\kappa B e^{(-\kappa h)} + G + \theta \frac{3kTr}{r_g^4}(h_{ct} - h)(h - r_g) \quad (S3)$$

And the potential of the microcapsule located at $h \leq h_{ct}$ can be integrated as:

$$\phi(h) = Be^{-\kappa h} + Gh + \theta \frac{3kTr}{2r_g^4}(h_{ct} - h)(h - r_g)^2 \quad (S4)$$

The relevant position is chosen at $h_{m'}$, which has potential energy of
$$\phi(h_{m'}) = Be^{-\kappa h_{m'}} + Gh_{m'} + \theta \frac{3kTr}{2r_g^4}(h_{ct} - h_{m'})(h_{m'} - r_g)^2 \quad (S5)$$

The relative potential energy can be calculated as: $\phi(h) - \phi(h_{m'})$, corresponding to eq. 8 in the main text.

For $h > h_{ct}$, the relative potential energy can be represented as following, equal to the eq. 10 in the main text:

$$\phi(h) - \phi(h_{m'}) = (Be^{-\kappa h} + Gh) - \left(Be^{-\kappa h_{m'}} + Gh_{m'} + \theta \frac{3kTr}{2r_g^4}(h_{ct} - h_{m'})(h_{m'} - r_g)^2\right) =$$

$$(Be^{-\kappa h} - Be^{-\kappa h_{m'}}) + (Gh - Gh_{m'}) - \theta \frac{3kTr}{2r_g^4}(h_{ct} - h_{m'})(h_{m'} - r_g)^2 =$$

$$B(e^{-\kappa h_{m'}}(e^{-\kappa(h - h_{m'})} - 1) + G(h - h_{m'}) - \theta \frac{3kTr}{2r_g^4}(h_{ct} - h_{m'})(h_{m'} - r_g)^2 \quad (S6)$$

For $h \leq h_{ct}$, the relative potential energy can be represented as following, equal to the eq. 11 in the main text:

$$\phi(h) - \phi(h_{m'}) = \left(Be^{-\kappa h} + Gh + \theta \frac{3kTr}{2r_g^4}(h_{ct} - h)(h - r_g)^2\right) - \left(Be^{-\kappa h_{m'}} + Gh_{m'} + \theta \frac{3kTr}{2r_g^4}(h_{ct} - h_{m'})(h_{m'} - r_g)^2\right) = \left(Be^{-\kappa h} - Be^{-\kappa h_{m'}}\right) + (Gh - Gh_{m'}) + \theta \frac{3kTr}{2r_g^4}[(h_{ct} - h)(h - r_g)^2 - (h_{ct} - h_{m'})(h_{m'} - r_g)^2] = B(e^{-\kappa h_{m'}}(e^{-\kappa(h - h_{m'})} - 1) + G(h - h_{m'}) + \theta \frac{3kTr}{2r_g^4}[(h_{ct} - h)(h - r_g)^2 - (h_{ct} - h_{m'})(h_{m'} - r_g)^2] \quad (S7)$$

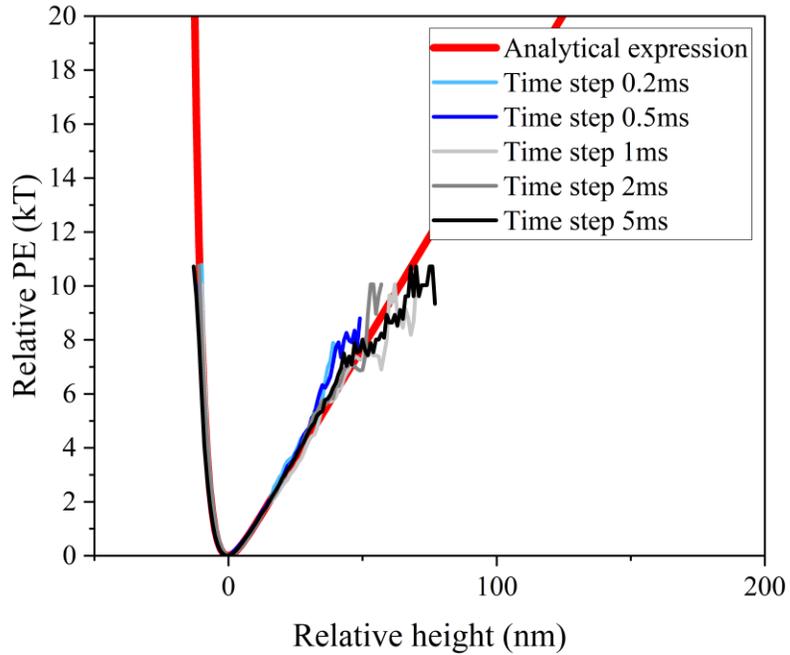

**Figure S1. Control simulation results were compared to analytical results to determine the appropriate simulation time step.**

| Time step | 5ms | 2ms | 1ms | 0.5ms | 0.2ms |
|---|---|---|---|---|---|
| $\chi^2$ | 15.34 | 2.80 | 5.80 | 5.25 | 4.75 |

**Table S1. Chi-square ($\chi^2$) between control simulation results and analytical expression.** Control Brownian dynamics simulations were conducted for a series of time steps ranging from 0.2 ms to 5 ms. The simulation accounted for gravity and electrostatic forces in particle-boundary interactions, as well as hydrodynamic restrictions near the boundary. The simulation results were compared with analytical expression and $\chi^2$ values were calculated. Based on the results, a time step of 0.2 ms was selected as the most appropriate for the subsequent simulations in the paper.

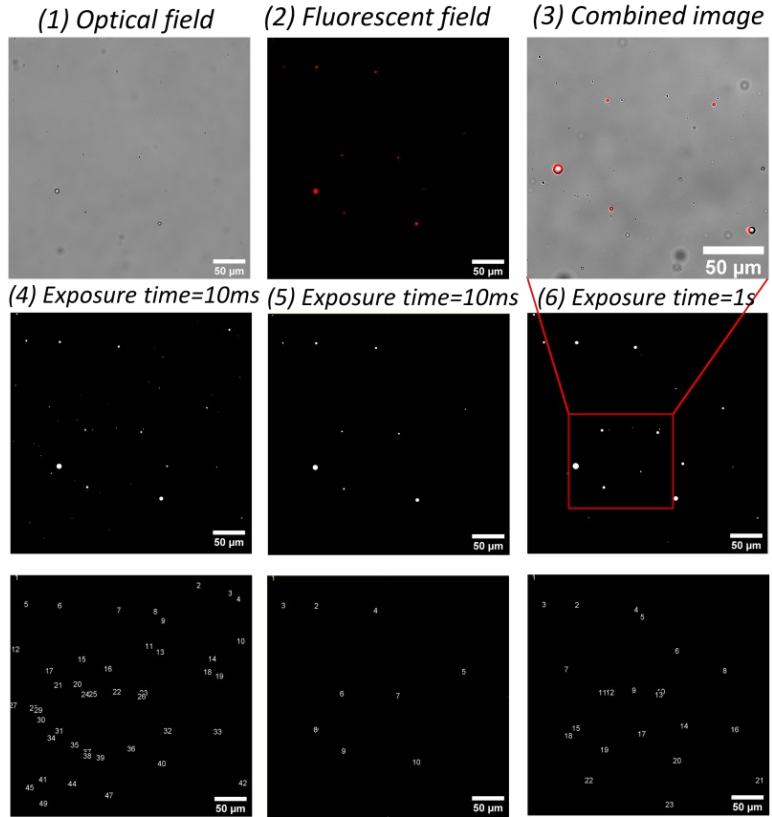

**Figure S2. Fluorescent measurement of microcapsule core-shell structure.** Fluorescent tag Nile Red was added in the oil core phase during fabrication process. The fluorescent microcapsules were adhered to the glass boundary, and optical images were taken to determine the outer diameter $D_{shell}$ of the microcapsule shell. Oil core sizes $D_{core}$ of the same fluorescent microcapsules were measured by the fluorescent microscopy. The shell thickness $\delta_{shell}$ was then calculated as $\delta_{shell} = D_{shell} - D_{core}$. In our measurements, $D_{shell} = 4.72 \pm 1.75\ \mu m$, $D_{core} = 43.372 \pm 1.61\ \mu m$, and $\delta_{shell} = 1.36 \pm 0.77\ \mu m$. Note here that optical imaging is not a quantitative method for measuring the shell

thickness. However, the distinct sizes of oil core and polymer shell provide the evidence of core-shells structure of the microcapsules.